# Role of heat and mechanical treatments in the fabrication of superconducting Ba$_{0.6}$K$_{0.4}$Fe$_2$As$_2$ *ex-situ* Powder-In-Tube tapes

A. Malagoli[1], E. Wiesenmayer[2], S. Marchner[2], D. Johrendt[2], A. Genovese[3,4], and M. Putti[1,5]

[1] CNR-SPIN, Corso Perrone 24, 16152 Genova, Italy
[2] Department Chemie, Ludwig-Maximilians-Universität, München, Germany
[3] King Abdullah University of Science and Technology, 23955-6900 Thuwal, Saudi Arabia
[4] Italian Institute of Technology, 16163 Genova, Italy
[5] Department of Physics, University of Genova, Via Dodecaneso 33, 16146 Genova, Italy

**Abstract** Among the recently discovered Fe-based superconducting compounds, the (K,Ba)Fe$_2$As$_2$ phase is attracting large interest within the scientific community interested in conductor developments. In fact, after some years of development, critical current densities $J_c$ of about 10$^5$ A/cm$^2$ at fields up to more than 10 T have been obtained in powder in tube (PIT) processed wires and tapes. Here we explore the crucial points in the wire/tape fabrication by means of the *ex-situ* PIT method. We focus on scaling up processes which are crucial for the industrial fabrication. We analyzed the effects on the microstructure of the different heat and mechanical treatments. By an extensive microstructural analysis correlated with the transport properties we addressed the issues concerning the phase purity, the internal porosity and crack formation in the superconducting core region. Our best conductors with a filling factor of about 30 % heat treated at 800 °C exhibited $T_c$ = 38 K the highest value measured in such kind of superconducting tape. The microstructure analysis shows clean and well connected grain boundaries but rather poor density: The measured $J_c$ of about 3·10$^4$ A/cm$^2$ in self-field is suppressed by less than a factor 7 at 7 T. Such not yet optimized $J_c$ values can be accounted for by the reduced density while the moderate in-field suppression and a rather high *n*-factor confirm the high homogeneity and uniformity of these tapes.

## 1. Introduction

The discovery of superconductivity in Fe-based compounds [1] has raised, during the last years, new enthusiasm into both fundamental and applicative researches [2, 3, 4, 5]. Among several families of superconducting materials, (*AE*)Fe$_2$As$_2$ compounds (*AE*= alkaline earth metals) represent the better candidates to develop practical conductors. The reasons lie on their particularly interesting properties: i) a critical temperature $T_c$ ~ 39K, ii) a very high and nearly isotropic upper critical field, iii) reduced thermal fluctuation and iv) a not so detrimental nature of grain boundaries [4, 6, 7, 8]. Such properties make this family the most suitable candidate for high field applications (> 20 T), as Nb-based superconductors are not able to reach these working ranges and high-T$_c$ cuprate superconductors have still difficulty to stand out. Concerning such applications, we need high critical current density $J_c$ and a weak dependence on the applied magnetic field. A level of 10$^5$ A/cm$^2$ at the field of interest is considered the minimum target to justify the use of a superconductor in industrial applications. The achievement of all these characteristics into a single wire represents a major technological breakthrough. Using the coated-conductor technology a $J_c$ of 1MA/cm$^2$ has been measured in Co-doped BaFe$_2$As$_2$ grown on Fe/IBAD-MgO-buffered metal tape [9], however the most explored technique for wire manufacturing is the Powder In Tube (PIT) method [10, 11,12]. Weiss *et al.* [13] showed how high pressure applied on the wire during the heat treatment at relatively low temperature results in fine grains and enhancement of the superconducting core density and thus of $J_c$. The effectiveness of the texturing induced by rolling in $J_c$ enhancement has been reported in ref.s[12, 14]. More recently, many papers described how a combination of densification and texturing processes realised through uniaxial pressing [15, 16] or hot pressing [17, 18] allow to achieve $J_c$ values of industrial interest. The best results so far, i.e. $10^5$ A/cm$^2$ at 10T, are reported by Lin et al.[19] on a hot pressed tape and Gao et al [20] on a stainless steel/Ag sheathed tape. These encouraging results push on the development of scalable processes, which keep or even improve transport properties over long length conductors. From this point of view, we can observe that the hot and uniaxial pressing methods are not easily transferable to long wires.

We are interested in exploring the actual potential of a simple PIT process applicable to few cm wires as well as to longer ones. We therefore focused on heat treatments performed without any additional pressure and on different possible deformation processes including the use of both single and double sheath. Here we report powder and tapes/wires preparation and a deep analysis of the chemical, structural and superconducting properties by means of X-ray diffraction (XRD), transmission electron microscopy (TEM), scanning electron microscopy (SEM), transport and magnetic measurements. Our best conductor shows the highest $T_c$ measured so far in such kind of superconducting tapes and a $J_c$ of about 3·10$^4$ A/cm$^2$ with a filling factor of about 30%. We will focus on the effects of the annealing temperature on the phase chemical properties and on the brittleness of the superconducting core, which appears to be a crucial point for the cold working.

## 2. Experimental

*2.1 Powder Synthesis*
The of Ba$_{0.6}$K$_{0.4}$Fe$_2$As$_2$ powders were synthesized by heating stoichiometric mixtures of the elements (purities > 99.9 %) in alumina crucibles sealed in silica tubes under an atmosphere of purified argon. In order to preserve as much potassium as possible from evaporation, alumina inlays were used. The mixtures were heated under a rate of 50 K/h up to 600 °C and held at this temperature for 15 h. The reaction products were homogenized, enclosed in a silica tube and annealed at 650 °C for 15 h. Afterwards, the mixtures were homogenized again, cold pressed into pellets and sintered for 20 h at 750 °C. After annealing, the samples were cooled down to room temperature by switching off the furnaces. Small fractions of the impurity phase FeAs (< 3.5 %) were detected due to potassium evaporation. Lattice parameters, Ba/K atomic ratio and the crystal purity were assessed by X-ray powder diffraction (Cu-K$_{\alpha 1}$-radiation) and Rietveld refinement using the TOPAS package [21].

*2.2 Tapes and wires preparation*
The so prepared powder was used to fill an Ag tube with an outer diameter of 8 mm and an inner diameter of 5 mm. The Ag for the sheath was chosen because of its ductility and chemical compatibility with the superconducting phase even at high temperature. This composite tube was cold worked through drawing, flat rolling and groove rolling to obtain wires or tapes with different thickness and different level of powder compaction. In order to further enhance the strength of the deformation, in one of the samples a double metallic sheath composed by an external Ni and an internal Ag tubes was used. Table 1 lists the samples analyzed in this work and specifies their preparation conditions. Each wire/tape length so obtained was about 2 m.

**Table 1.** Geometry of studied conductors and their preparation routes

| Sample (sheath) | Working | Size |
|---|---|---|
| A  (Ni/Ag) | groove rolled wire | 1 x 1 mm$^2$ |
| B  (Ag) | drawn wire | Ø 0.9 mm |
| C  (Ag) | drawn + flat rolled tape | 0.5 mm thick |
| D  (Ag) | drawn + flat rolled tape | 0.4 mm thick |
| E  (Ag) | groove rolled wire | 0.9 x 0.9 mm$^2$ |
| F  (Ag) | drawn + groove rolled wire | 0.9 x 0.9 mm$^2$ |
| G  (Ag) | groove rolled + flat rolled tape | 0.4 mm thick |

In figure 1 the cross sections for the samples A, B, D, E are shown.

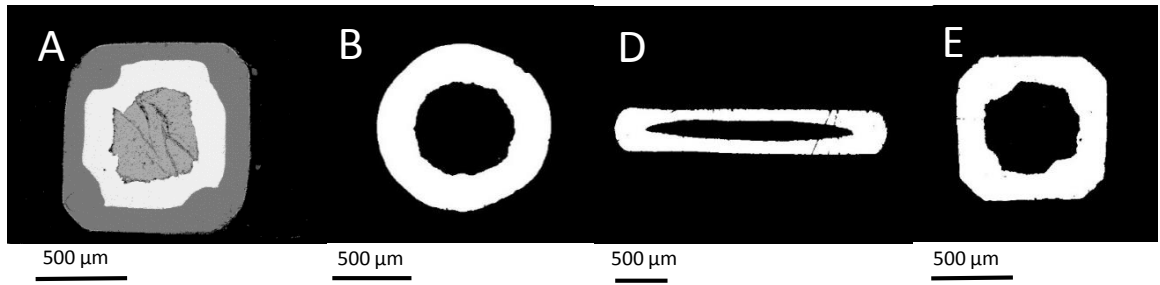

**Figure 1:** Cross sections of samples A,B,D and E.

10 cm long samples underwent a sintering heat treatment in 1 bar flowing Ar in a three-zone tubular furnace with a homogeneity zone (± 0.5 °C) of 16 cm at different temperatures between 700 – 850 °C. Before the heat treatment, the ends of all samples were sealed by dipping them into molten Silver. This sealing should prevent inner gas or element escape.

### 2.3 Transport and magnetic characterization
The resistivity measurements on the wires/tapes were performed with a DC-four-probe system with variable temperature from 300 to 4.2 K. Short sample pieces of about 6 mm in length were employed for the magnetization measurements vs. temperature, performed with a commercial 5.5 T MPMS Quantum Design Squid magnetometer using a background field of 10 Oe. Critical current $I_c$ measurements were performed by a DC-four-probe home-made system in liquid helium bath (4.2 K) and magnetic field up to 7 T.

### 2.4 Electron Microscopy analysis
Some Ba$_{0.6}$K$_{0.4}$Fe$_2$As$_2$ conductor samples were analyzed *via* transmission electron microscopy (TEM) and scanning electron microscopy (SEM). The samples were prepared in two steps, by mechanical milling and subsequent ion thinning with a Gatan precision ion polishing system (PIPS). High resolution TEM (HRTEM), energy filtered TEM (EFTEM) and high angle annular dark field (HAADF) scanning TEM (STEM) measurements were performed by a FEI Titan Cube microscope, working at 300 kV, equipped with a Schottky electron source, a CEOS spherical aberration corrector of the objective lens, which allows to reach a sub-angstrom resolution (0.9 Å), and a post column Gatan Image Filter (GIF) Tridiem . Spatially resolved chemical analysis was performed in STEM mode via energy dispersive X-ray spectroscopy (EDX) using a EDAX Silicon(Li) detector with an area of 30 mm$^2$ and chemical quantification calculated using the standardless Cliff-Lorimer method. TEM characterizations were carried out using a double tilt holder equipped with a beryllium tip to align correctly the lamellae along zone axes and to reduce the background in EDX analysis.

SEM characterization of as-obtained polished sections was performed using a Jeol JSM-7500F scanning electron microscopy equipped with a cold field emission gun. SEM analysis was carried out with an accelerating voltage of 10 kV and using a Rutherford backscattered electron imaging (RBEI) detector to increase electron contrast by adding a compositional contribution. SEM-EDX chemical analysis was performed using an Oxford X-Max 80 system with a SDD detector of 80 mm$^2$.

## 3. Results

### 3.1 Powder

A powder diffraction pattern and the Rietveld fit of one sample free from impurity phases is depicted in figure 2a while the X-ray powder diffraction patterns of seven powder batches are shown in figure 2b.

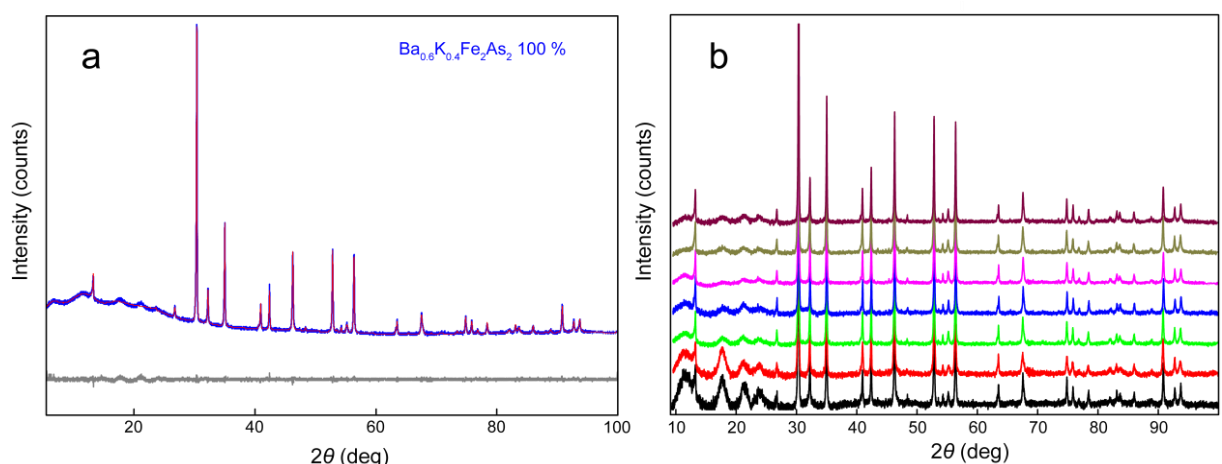

**Figure 2: a)** X-ray powder diffraction pattern (blue) and Rietveld fit (red) of one of the $Ba_{0.6}K_{0.4}Fe_2As_2$ portions. **b)** X-ray powder diffraction patterns of the seven different samples of $Ba_{0.6}K_{0.4}Fe_2As_2$ that were used in the wire fabrication process.

The three broaden reflections between 16 and 26 °$2\theta$ are a result of measurement artifacts. In order to obtain high quality material with a homogeneous potassium distribution we emphasize that the samples have to be homogenized accurately after each step. Magnetization and superconducting transition temperatures of the finely ground powder samples were measured between 3.5 K and 50 K and are shown in figure 3 normalized to the value measure at the lowest temperature. All batches show the same transition temperature ($T_c$ = 38.5K) and a narrow superconducting transition. All these results demonstrate the high reproducibility and reliability of the powder preparation process, a necessary condition for a possible scaling up.

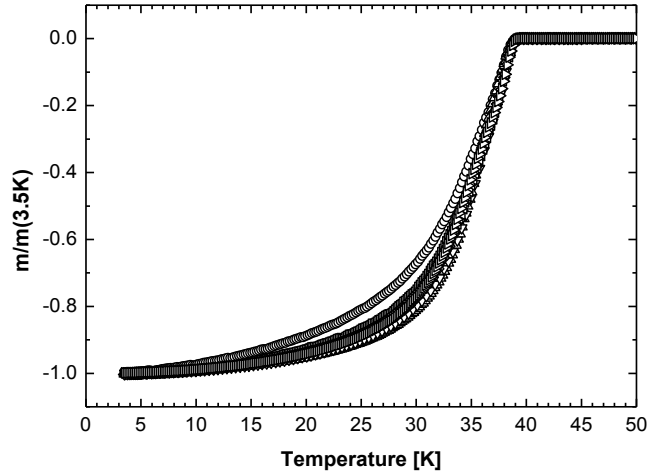

**Figure 3:** normalized magnetization of the seven batches of $Ba_{0.6}K_{0.4}Fe_2As_2$: all batches show the same transition temperature $T_c$= 38.5K

*3.2 Tapes and Wires*

*3.2.1 Effects of the heat treatment temperature*

In order to optimize the sintering temperature the sample D was heat treated at three different temperatures: 700 °C, 800 °C and 850 °C. Figure 4 shows the measured resistance (left panel) and susceptibility (right panel) superconducting transition. In the left panel $T_c$ corresponding to zero resistance of 36.8 K and 36.6 K were measured for the samples treated at 800 °C and 850 °C respectively, while for the sample treated at 700 °C we observed a drop down to about 33 K. Concerning the shape of the transitions we can observe that the sharpest one is for the 800 °C sample denoting a higher homogeneity of the superconducting core.

The magnetic susceptibility measurements reported in the right panel, exhibit similar $T_c$ values for 800 °C and 850 °C samples, being 37.7 K and 37.0 K, respectively. This is the highest measured $T_c$ in tapes and wires so far. However, the behavior of the three examined susceptibilities is quite different. Indeed, while for the 800 °C sample an almost complete shielding and a narrow superconducting transition are observed, for the 850 °C and 700 °C samples only a partial shielding and broader transition are measured. In particular, the 700 °C sample shows an almost negligible shielding. The overall behavior of the 700 °C annealed sample indicates that this temperature is too low to recover the defects induced by the mechanical deformation and to improve the grain connectivity, as it results -in its degraded superconducting properties.

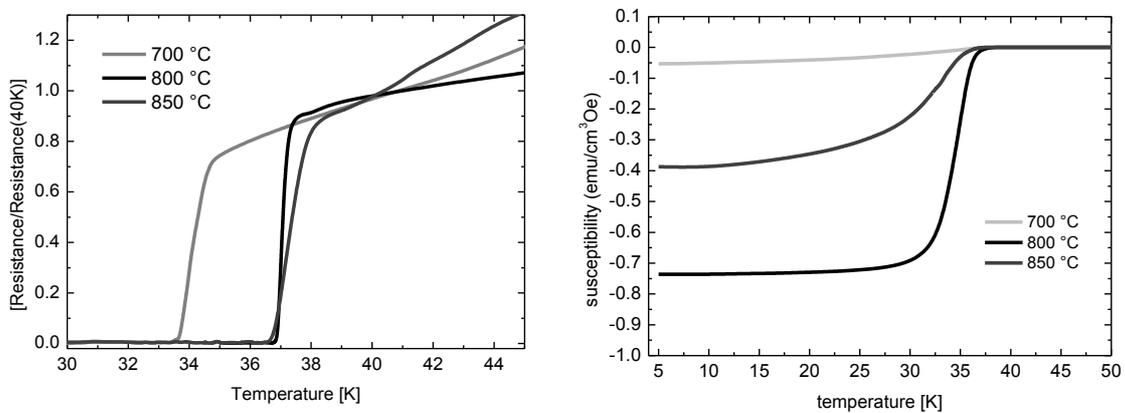

**Figure 4:** Left panel: Resistance measurement of samples annealed at 700 °C, 800 °C and 850 °C. Right panel: susceptibility measurement at 700 °C, 800 °C and 850 °C.

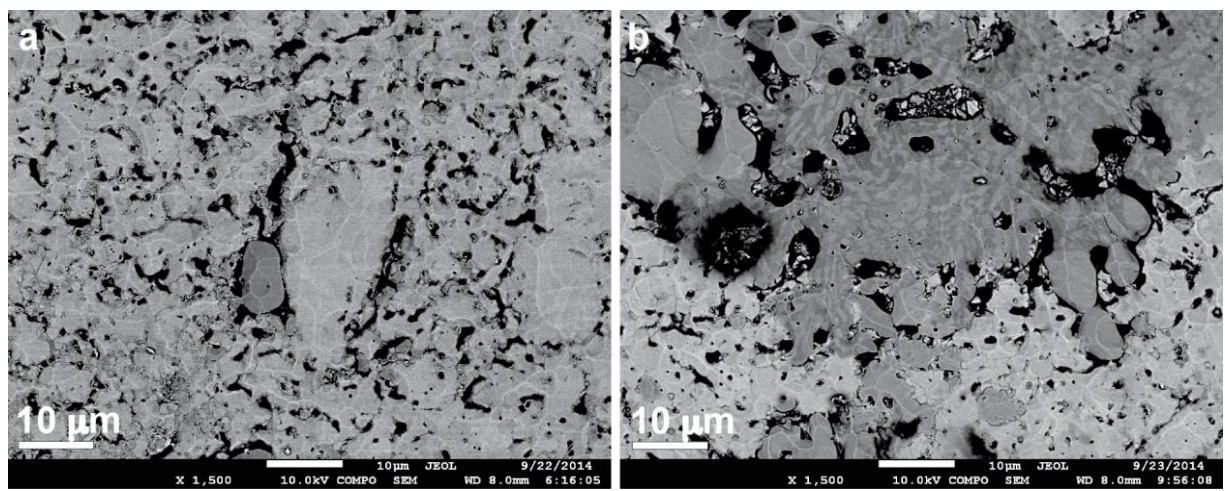

**Figure 5**: HR-SEM images, acquired with RBEI detector, of a tape 0.4 mm treated at 800 °C (left) and 850 °C (right). The electron contrast changes are directly correlated to phase variations, lighter depicts $Ba_{0.6}K_{0.4}Fe_2As_2$ and darker $Fe_xAs_y$. The black areas correspond to voids within the tapes.

Therefore we focused our attention on the samples treated at 800 °C and 850 °C .
Figure 5 shows high resolution SEM (HR-SEM) images of the core regions of samples treated at 800 °C (left panel) and 850 °C (right panel), respectively. The electron contrast, depicted as grey scale, revealed clearly chemical differences within the tapes, where the light grey regions correspond to the pure superconducting phase, while the darker areas to $Fe_2As$ and $FeAs$ secondary phases. It is therefore, evident how an incremental temperature of only 50 °C is crucial for the nucleation and growth processes of $Fe_xAs_y$ secondary phases. In particular, after the thermal treatment at 850 °C the tapes exhibited large $Fe_xAs_y$ domains grown between the superconducting grains. Differently, in the sample treated at 800 °C, the secondary phases were negligible and generally formed isolated grains of small size. A common feature present in all the images is a large amount of voids, which testify the poor density of the tapes.
Detailed EFTEM, HRTEM and STEM characterizations were performed in order to discriminate textural and chemical features of both the superconducting and secondary phases and their relationship at grain boundary interfaces. In figure 6 we reported the results of these analyses carried out on a representative sample treated at 800 °C. EFTEM observations, obtained using an energy slit 10 eV wide just to increase electron contrast by removing the inelastic electron scattering, displayed superconducting crystal grains with dislocation fields arranged at high angle in respect to their grain boundaries, whose interfaces exhibited straight profiles almost parallel to the tape rolling-direction (figure 6a). HAADF STEM imaging and STEM-EDX chemical analysis exhibited no variations in electron contrast and no chemical differentiation, confirming a homogeneous stoichiometry consistent with the only one superconducting phase, as clearly displayed by both linear and areal STEM-EDX mapping (figures 6b and c). HRTEM observations revealed the fine structures of grains and their interfaces (figure 6d). The superconducting grains exhibited no lattice distortion or change in lattice periodicity across them, but only semi-coherent and incoherent interfaces at grain boundaries. In particular, proximal grains displayed slight misalignments with similar crystallographic orientations, as clearly inferred from their corresponding HRTEM and fast Fourier transform (FT) analysis where direct and reciprocal vector relations are shown (Figure 6d, e).

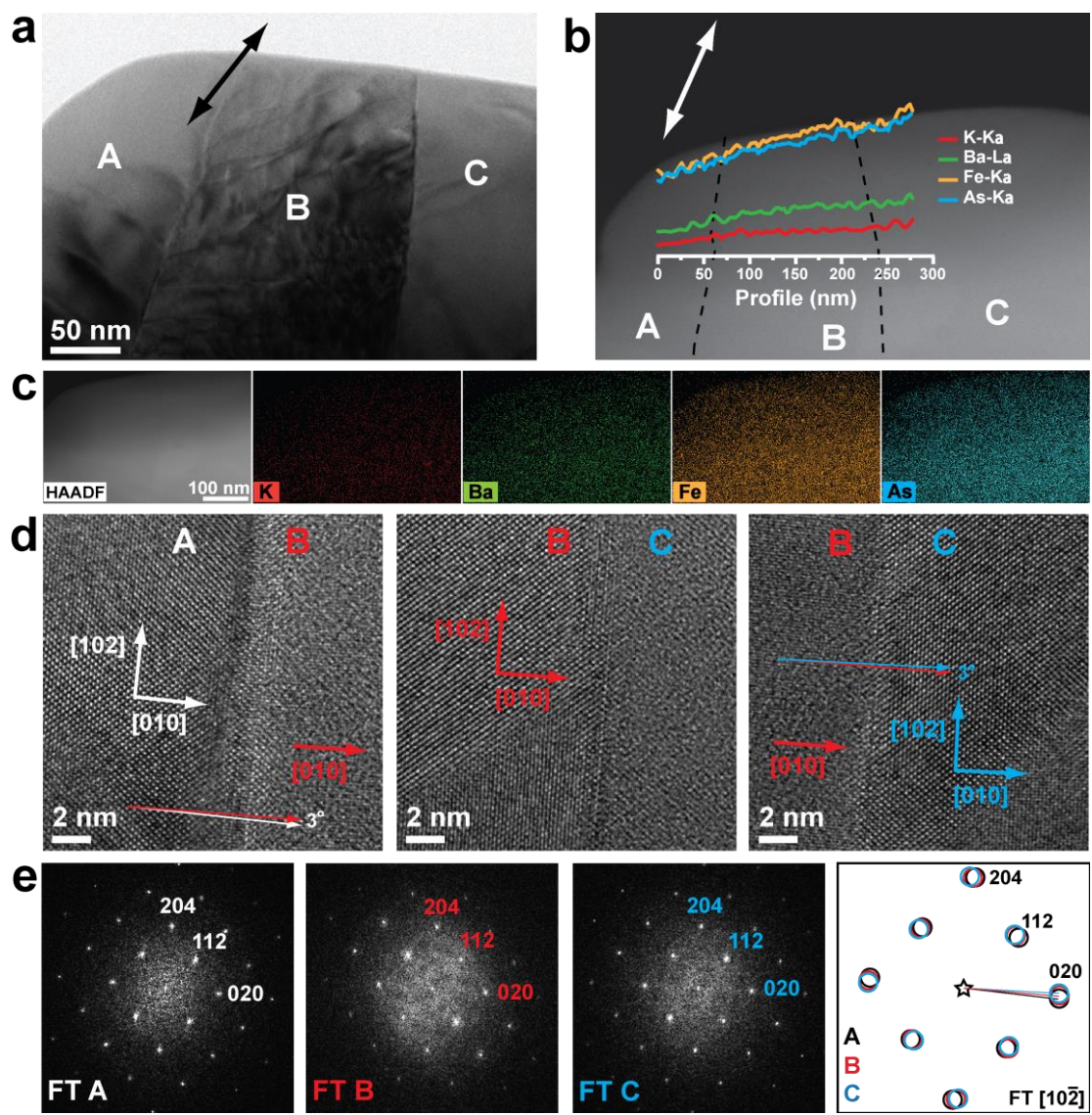

**Figure 6:** TEM study of the core region of a tape sample treated at 800 °C. **a)** EFTEM displaying three grains (A, B, C) with almost parallel interfaces; black arrow indicates the tape rolling-direction. **b)** HAADF image and STEM-EDX linear map revealing no chemical changes, the slight intensity increasing of element signals from left to right is due to the small thickness increase along the profile direction; black dotted lines mark grain boundaries, white arrow the tape rolling-direction. **c)** HAADF image and STEM-EDX areal map revealing homogeneous stoichiometry. **d)** HRTEM of A, B and C grains observed along the common $[\bar{2}01]$ zone axis and exhibiting rotational misalignments of 3°. **e)** Corresponding Fourier transforms (FT) of the grains revealing vector relationships consistent with a tetragonal symmetry and showing typical 204, 112 and 020 reciprocal spots of $Ba_{0.6}K_{0.4}Fe_2As_2$ phase (space group $I4/mmm$).

STEM-EDX and HRTEM characterizations carried out on the sample treated at 850 °C showed chemical and structural heterogeneities which might be correlated to decomposition reactions occurring at high temperature (figure 7). HAADF STEM-EDX study displayed evident compositional variations consistent with the presence of superconducting and non-superconducting phases. In particular, STEM-EDX chemical mapping of core regions highlighted a homogeneous distribution of As and a heterogeneous distribution of metallic cations K, Ba and Fe across the crystal grains, whose interfaces displayed curving profiles (figure 7a). These features might be ascribed to a process of preferential depletion and diffusion of metallic cations that led to chemical and structural modifications at local scale. In particular, K and Ba (elements of I and II group respectively) have similar electron affinity and electronegativity parameters, while Fe (VIII group) shows different values. These characteristics together with higher thermal diffusion regime, would have allowed the nucleation and growth of secondary $Fe_xAs_y$ phases within the superconducting phase. HRTEM observations clearly revealed the crystalline nature of these (K, Ba)-depleted secondary phases nucleated at grain boundaries and characterized by semicoherent interfaces (figure 7b).

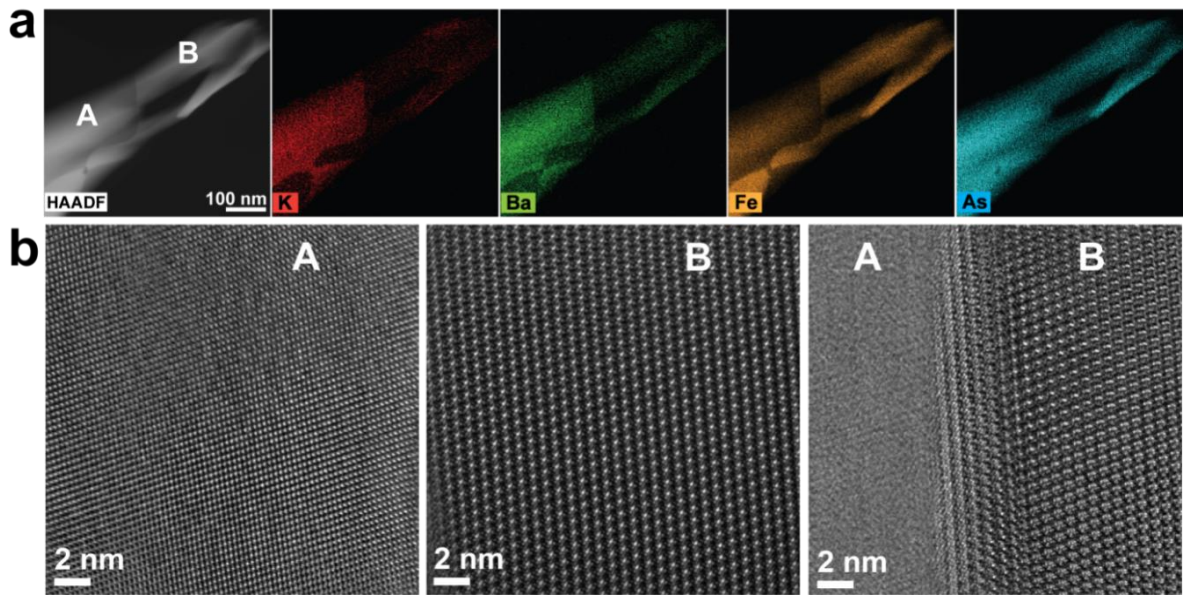

**Figure 7:** TEM study on core region of a wire sample treated at 850 °C. **a)** HAADF image and STEM-EDX areal map revealing the phase decomposition process at local scale (A superconducting and B non-superconducting phase, respectively); STEM-EDX chemical maps showing the elements distribution: homogeneous for As; heterogeneous for metallic cations. **b)** HRTEM of A, B grains displaying their different crystal structure and grain boundaries characterized by a semicoherent interfaces.

*3.2.2 Effect of the deformation*
The effects of the different deformation processes on the microstructures and on the transport properties were investigated in order to improve the superconducting performances.
As to compare extreme deformation conditions, we considered two samples: the sample A has a double metallic sheath, composed by Ni and Ag, and it was deformed by groove rolling: this means that it underwent the strongest deformation among all samples. The sample B has a single Ag sheath and was deformed just by drawing, thus we can assume that it underwent the softest deformation.
To remove the sheath and expose the superconducting core, avoiding as much as possible artificial damage or crack that can occur, for example, by polishing procedure, 12 mm long samples underwent a chemical etching. In particular we used a solution based on ammonium hydroxide (50 vol% aqueous solution) and hydrogen peroxide (30 wt% aqueous solution), mixed in equal amounts and partly diluted by distilled water. Figure 8 shows the longitudinal cross section of the samples A (upper picture) and B (lower picture) without any heat treatment. In the pictures just few but representative millimeter sample lengths are shown. It is evident how the microstructures of the two superconducting filaments are different. In the sample A we can observe a denser core structure but also several transversal cracks, obstacles to the super-current flow, and a not uniform filament shape. Differently, the sample B looks uniform without evidence of transversal cracks but the presence of a distributed porosity indicates a not so high core density. Actually several longitudinal lines are evident, typical of a cold working where the longitudinal component of the mechanical effort is anything but negligible as in drawing. However, because these defects are in the same direction as the current flow they are not an obstacle.

Then we analyzed the microstructures of the samples E and G that underwent an intermediate cold working, in which a combination of stronger deformation (groove-rolling and flat-rolling) with softer sheath (Ag) was performed.

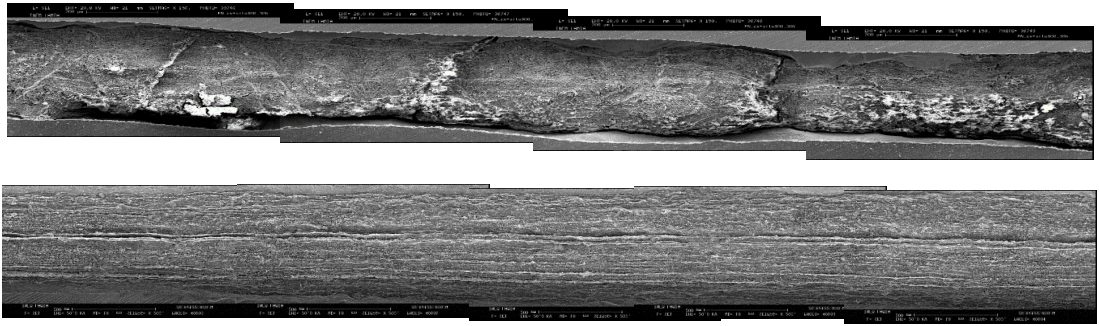

**Figure 8.** Longitudinal cross-sections of sample A (double Ni + Ag sheath) deformed by groove rolling (upper picture), and sample B (single Ag sheath) deformed by drawing (lower picture).

Figure 9 shows the longitudinal cross section of the wire E, at an intermediate stage - (a) and at the final stage of the deformation (b). At the intermediate stage, the superconducting filament looks uniform even if a visible porosity is still present. In the final wire, the shape of the core is still uniform but several cracks, indicated by the arrows, appear in transversal direction. However, it is evident as well that the filament density is largely enhanced, confirming the high potential of the groove rolling process in compacting the powders inside the wire. In figure 9 c, the image of the same sample E after the heat treatment at 800 °C is reported. The effects of the sintering process is to enhance the size of the grains and their connectivity. In addition, it is also possible to recover most of the cracks caused by the cold working although, as highlighted by the circles, some of them remain. The same results are observed for the sample G (figure 9 d) obtained from sample E by rolling it to a tape: even after the heat treatment, some cracks are still evident in the microstructure.

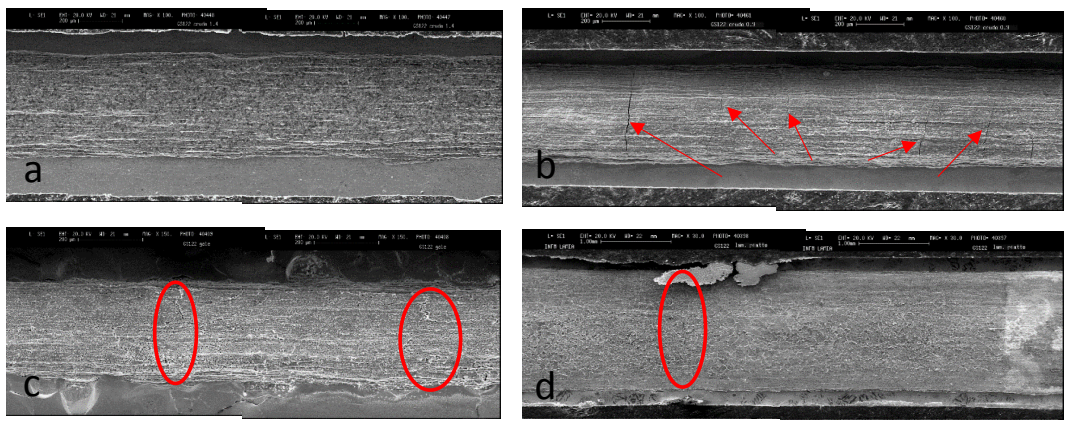

**Figure 9.** Longitudinal cross sections for sample E at an intermediate stage (a) and at the final stage of the groove-rolling process (b), and after the heat treatment (c); for sample G after the heat treatment (d).

*3.2.3 Critical Current Density*

The transport properties of all the samples after the heat treatment at 800 °C and 850 °C were investigated. For the samples A, E and G, which presented high density of transversal microcracks, no appreciable critical currents were measured. For the samples B, C, D and F the resulting critical current densities $J_c$ after the heat treatment at 850 °C were almost negligible, while after the heat treatment at 800 °C the $J_c$ as a function of the applied magnetic field are reported in figure 10.

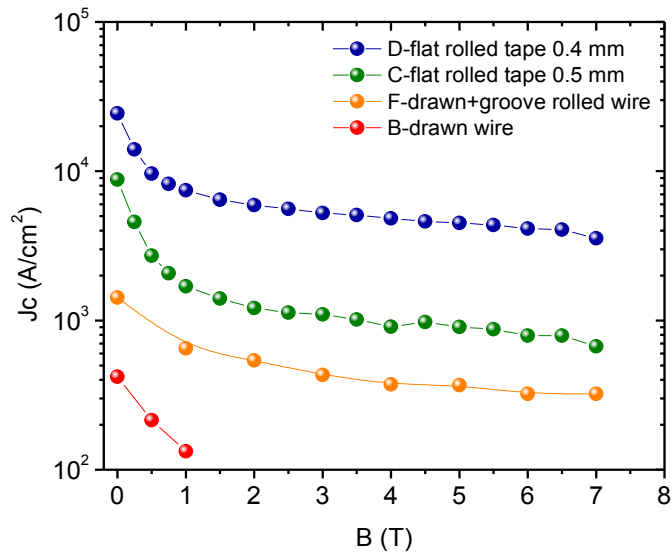

**Figure 10.** Critical current density vs applied magnetic field for the samples B, C, D and F.

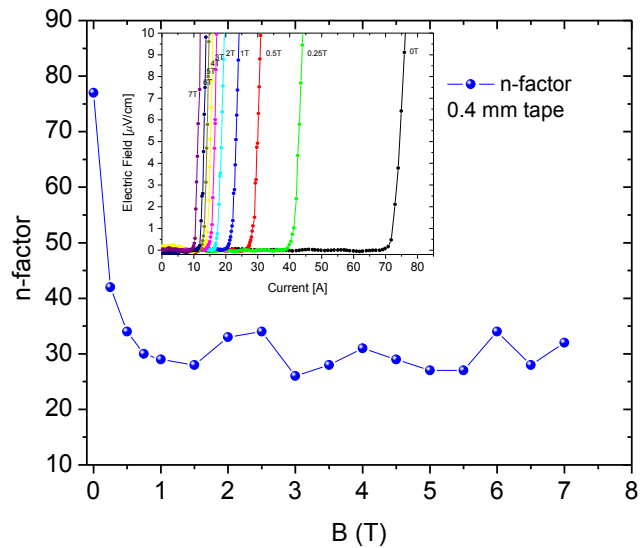

**Figure 11.** n-factor vs applied magnetic field of the sample D evaluated from the measured V-I curve reported in the inset.

With reference to table 1 the samples B, C, D and F were prepared starting with the drawing process: B is an as drawn wire, C and D (both tapes) were first drawn into wires of 1.4 mm in diameter and later flat-rolled in several steps down to 0.5 and 0.4 mm in thickness respectively; finally the sample F was first drawn and later groove-rolled in just few final steps to obtain a square 0.9 x 0.9 mm wire. $J_c$ increases as the strength of the cold working increases, i.e. as the powder compaction is enhanced. The behavior of $J_c$ in field is similar for all samples and consistent with the known literature for this compound, described by an initial drop (within 2 T) and a following almost flat trend at higher field. It is worth to note that in-field suppression is rather limited (less than a factor 7 between the self-field values and the value at 7 T). The highest $J_c$ values were obtained for the sample D. From the measured *V-I* curve we extracted the *n*-factor reported in figure 11. Except for the very high values at low fields, we can observe an average value of 30, independent of the field. This result is a strong indication of the high homogeneity and uniformity of this tape.

### 4. Discussion

The main aim of this work was to explore the crucial steps in wire/tape fabrication by mean of the *ex-situ* PIT method for optimally potassium doped Ba122. We considered the fabrication steps crucial from the point of view of a scaling up of the process to an industrial level. Given the importance of using pure starting materials, we preferred to investigate the effect of temperature and mechanical treatment on optimized powders without to pay particular attention in making scalable the process for powder production. Outstanding results in short time have been already obtained [19, 20] on short PIT processed samples, making even more important and urgent to find a reproducible and reliable way to fabricate 122 superconducting wires in lengths useful for real applications. Discussing the results described in this work, we focus on the main preparation stages that should be considered.

Concerning the sintering condition, several processes were employed so far, such as hot isostatic pressure [13] at 600 °C or hot uniaxial pressure [19] at about 900 °C. Gao Z. *et al.*, put the samples into a stainless steel tube sealed by arc welding in Ar atmosphere before the heat treatment [20] reaching a pressure of about 3-4 bar at 850 °C. Additional pressure produces the enhancement of the core density and can prevent element losses, leading to improved superconducting properties but it is clear that the employed methods are not easily transferable to long length wires.

We used a sintering heat treatment in 1 bar flowing Ar. No additional pressure was applied on the sample and to prevent elements and gas escape and, as a consequence, a decreasing of the superconducting performances, we sealed the ends of the samples. This process is straightforward scalable to industry being the same used, for example, for $MgB_2$ manufacturing [22].

In this contest, it is interesting to compare the different temperatures used for the heat treatment. The heat treatment temperatures reported for the best conductors which underwent additional pressure [19, 20] are of 900 °C and 850 °C. Our best samples are heat treated at 800 °C. These samples exhibit optimal critical temperature and quite pure superconducting phase, with just a negligible amount of secondary phase probably coming from the starting powders. With applying temperatures higher than 800 °C a process of depletion and diffusion of metallic cations occurs, which causes the nucleation and growth of secondary $Fe_xAs_y$ phases from the superconducting phase as clearly revealed by HRTEM observations; the detrimental effects on the global current of $Fe_xAs_y$ wetting phases at the grain boundary is well known [23, 24]. We can speculate that the additional pressure limits the depletion and diffusion of elements such as Ba and K and, consequently, inhibits the formation of $Fe_xAs_y$ secondary phases allowing reaching higher temperatures. Thus, we conclude that heat treatment temperature should be optimized as a function of the applied pressure, but it is not a parameter that uniquely determines the sample performance. Indeed, higher heat treatment temperature could improve the grain sintering, but also increase the grain size. This latter aspect is not advantageous for enhancing the critical current, as discussed by Weiss *et al.* [13] who developed fine grain in $(Ba_{0.6}K_{0.4})Fe_2As_2$ wires sintered at 600 °C in order to increase grain boundary density.

The second issue we faced is the cold deformation. We analysed the effects of different processes, i.e.

drawing and groove-rolling, on the microstructure. We observed that to get high core density a stronger deformation such as rolling is preferable, as already demonstrated for other PIT processed superconductors [25]. Nevertheless, we also observed that such a cold working provokes transversal cracks, limiting the supercurrent flow. This study has highlighted how this compound is more brittle then other technical superconductors, which are processed by PIT method, such as $MgB_2$ or BSCCO. However, the heat treatment at high temperatures is able to reduce this issue as seen in figure 9. This observation makes the role of intermediate heat treatments very important, since, if performed at appropriate steps of the deformation, they can almost totally recover the cracks formation. This could complicate the process, especially for multifilamentary conductors with finer filaments but still it seems to be the more promising path to pursue [16].

Finally, we discuss the transport measurements reported in fig.s 10 and 11. As already observed by other groups, $J_c$ increases as the strength of the deformation and thus the core density of the tape increases. Furthermore, increasing flat rolling steps can improve the grain texturing and thus contribute to enhancing $J_c$. In particular, higher density and improved texturing can explain such a better performance of the sample D with respect to the sample C. However, also in the most deformed sample (flat rolled 0.4 mm tape) the density is still quite low, as shown in the images of fig. 5. We suggest that this is the main reason for the not yet optimized $J_c$ values nearly one order of magnitude less that the best values reported in literature. Indeed the TEM analysis reveals clean grain boundaries (homogeneous stoichiometry across the grain boundaries) with induced texturing which should contribute in enhancing the superconducting performances. Moreover, the *n*-factor value, which is an indicator of the homogeneity of the superconductor, on a cm-scale sample, is around 30, almost independent on the applied magnetic field, confirming the optimal properties of our sample. In order to improve the density without applying any additional pressure an optimization of the deformation process seems to be needed. This can be obtained by employing composite and harder metallic sheaths to make the deformation stronger and thus enhancing the powder compression and by intermediate annealing treatments to recover the cracks.

## 5. Conclusions

In order to develop scaling up processes, which are crucial for the industrial fabrication, in this paper we explored some of, the crucial points in the fabrication of K doped Ba122 conductors by means of the ex-situ PIT method.

In particular, through an extensive microstructural analysis correlated with the transport properties we investigated the effects of different temperature treatment on the phase purity and grain sintering and mechanical treatments in the densification and crack formation.

We performed heat treatments at 700 °C, 800 °C and 850 °C without any additional external pressure. The best condition were found at 800 °C for which a pure superconducting phase and clean grain boundaries were observed. Indeed temperatures higher than 800 °C led to the formation of a secondary non-superconducting $Fe_xAs_y$ phases at the grain surface.

We varied the strength of the cold deformation both by applying different mechanical treatments (groove rolling and drawing) and by introducing harder external sheath. Starting from the softest deformation treatments $J_c$ increases substantially, indicating the beneficial effect of core densification. However, the strongest deformation processes produce cracks within the superconducting core strongly detrimental to the current transport. Thus, we pointed out that the brittleness of the superconducting phase is a crucial point that should be carefully considered.

Finally, aiming at a process that avoids the application of additional pressure, which is not an easily scalable process, we suggest to enhance the powder compaction by employing composite and harder metallic sheaths in combination with a strong deformation alternated with intermediate annealing treatments to recover the cracks.

**Acknowledgments**
This work has been supported by FP7 European project SUPER-IRON (grant agreement No.283204) and "Compagnia di SanPaolo". We thank Dr. Alice Scarpellini and Dr. Cristina Bernini for the help on scanning electron microscopy images acquisition.

13